# The gas equation for stream

S.L. Arsenjev, I.B. Lozovitski[1], Y.P.Sirik

*Physical-Technical Group*
*Dobroljubova street 2, 29, Pavlograd, Dnepropetrovsk region, 51400 Ukraine*

The equation of state for a gas stream as analog of well-known equation of state of gas medium in the static conditions is submitted. The connection of parameters of state and motion, and also the volume-flow rate and the mass-flow rate of the gas stream during change of its temperature and pressure as the result of external action is shown on example of modification of the represented equation.

The history of development of gas equation well-known by B.Clapeyron and D.Mendeleyev names includes more than the 200-years period (1661- 1874). However, the deep sense and significance of this law can be understood only now in the end of the XX century under the solution one of the classical problems of physics - development of the apparatus for description and quantitative evaluation of state and motion of gas stream and for simulation of fluid motion generally. This law which generalizes the special gas laws was determined methodologically in the context of the continual consideration of gas medium and phenomenologically - on the basis of direct experiment. Thus the level of generalizing is determined in not a smaller degree also by reflection in it of the mechanical equivalent of heat in the form of the gas constant (universal or specific) and connection with quantity of the substance.

It is remarkably that three special gas laws – R.Boyle and E.Mariotte, J.Charles, J.Charles and J.Gay-Lussac – was determined historically in sequence from a simple to a complex.

The importance of the first gas law established by R.Boyle (1661) and E.Mariotte (1676): $p_1V_1 = p_2V_2$ or $pV = const$, along with quantitative expression, consists in fundamentality of the approach, associated with concept of a compressibility as the property of bodies and mediums to change its volume and form under loading and elasticity as ability to recover initial volume and form at unloading. Besides, the conditions of holding of experiment ($T = const$) ensured the greatest universality of the approach to the test object. As Boyle wrote himself: «Both Pascal's experiments and experiments of our English friend (R.Hooke) were demonstrated that than greater weight acts to the air, all the more strong becomes its tendency to expansion and, therefore, its resisting strength (similarly to spring plates, which one become more strong, when them incurvate by the larger loads).» The approach is one both to the air, and to the spring plates: $T=const$. The fundamentality of Boyle's conceptual system, including such categories as compressibility, elasticity, undoubtedly, has formed the basis for determination of Hooke's law of an elasticity conformably to a tension of solid bodies.

The first law, as well as both following special gas laws, was determined at experiment with the fixed quantity of the gas placed in vessel hermetically closed by piston. In this connection, all three special gas laws are applicable to so-called closed-loop systems with fixed amount of gas medium.

The Boyle experiment conducted under stationary temperature characterizes the isothermal process of the pressure changing and volume of the given amount of gas. The isothermal change in modern thermodynamic interpretation is implemented at heating up of gas with a simultaneous external action (its expansion), at which one all heat injected toward gas is compensated by an expansion work ensuring an invariability of the reference temperature of the gas and its internal energy.

---

[1] Phone: (38 05632) 38892, 40596
E-mail: loz@inbox.ru

Later 15 years after Boyle's experiments, the first special gas law was established by Mariotte on the basis of self-maintained vast investigations predominantly of gas medium (though the Mariotte's formula for an estimate of stresses in a thin-wall cylindrical shell under action of pressure of fluid medium also occupies the substantial place in the strength theory.)

The second special gas law was established by J.Charles in 1787 and has two forms of writing: $p_2 = p_1(1 + \alpha_p \cdot t_2)$, $p_1/T_1 = p_2/T_2$ or $p/T = const$, it expresses the one aspect: in the closed vessel of invariable volume the heating up of gas leads to pressure increase so, that the ratio of the pressures is equal to the ratio of temperatures. The Charles's law, as well as the first special gas law, is an evidence of reversibility of behavior of gas, that is of its elasticity.

Process, which realizes in gas medium under invariable volume, is called the isochoric process in modern thermodynamic interpretation. The constancy not only volume of gas, but also its amount and, ultimately, of its weight density is typical for such process. All added heat is expended on heating of gas in this process that is on increase of its internal energy.

The third special gas law was established, as well as first law, twice and with the same differnce in years: in 1787 by J.Charles (is not published in time) and later 15 years, in 1802 by J.Gay-Lussac and has two forms of writing similarly to the second law: $V_2 = V_1(1 + \alpha_V \cdot t_2)$, where $\alpha_V = 1/273$ K$^{-1}$ = $\alpha_p$, and $V_1/T_1 = V_2/T_2$ or $V/T = const$.

This law implies, that for given and invariable amount of gas in the vessel, the volume of which can be changed, the heating up of gas leads to increase of its volume so, that the ratio of volumes is equal to the ratio of temperatures under constant pressure. Here again, the gas medium behaves elastically, as in two previous laws. It is typical for isobaric process in modern thermodynamic interpretation, that at heating up of invariable amount of gas is possible to ensure constancy of pressure in it by the resolution of free (thermal) expansion.

Gay-Lussac has attempted to unify the special gas laws in one generalized, taking into account the boundedness of effective parameters ($T, p, V$) and the unity of the tested medium. In result in 1826, he has received expression: $p \cdot V/T = const$. It is attested, that the phenomenological character of the special laws does not contain conceptual component permitting to uncover sense and magnitude of constant in right member of the equation.

The following attempt of a deduction of a generalized gas law was belonged by B.Clapeyron. In 1839 he has received $pV = BT$ and in 1840: $p \cdot V/T = BM$, where $B$ – proportionality constant, $M$ - weight of gas.

The opportunities of deriving of the generalized gas law by results of experiments with gas medium and by means of phenomenological approaches to problem were exhausted.

In the beginning of the second half of the XVIII century, J.Black has clearly explained the law of thermal equilibrium in parallel with the above mentioned investigations. In 1760, he has introduced the notion of heat capacity of bodies and mediums. To 1813 F.Delaroche and J.Berard have realized the first precise measurements of specific heats of gases. The results of these experiments boosted development of ideas about a possibility to make the work by gas at the expense of its internal energy at absence of thermal action from the outside. In 1816 P.Laplace, comparing heat capacities of isochoric by Charles's and isobaric by Charles and Gay-Lussac's processes, has determined their inequality that in isobaric process the heat capacity of gas is bigger in $k = c_p/c_V$ time. In result, in 1816 Laplace has resulted to the final form the formula for speed of sound in gas medium, that is from form: $a = \sqrt{p/\rho}$ previously deduced by I.Newton on the basis of the Boyle's law and E.Torricelli's formula in form: $a = \sqrt{k \cdot p/\rho}$.

The invention of the pneumatic fire in France in 1803, Gay-Lussac's experiments by compression and expansion, and also transflow of gas in the heat-insulated vessels have resulted to the concept of adiabatic process in 1807 and linkage of thermal energy with work of the expansion and compression of gas. In result, in 1823 S.Poisson has deduced the equation of adiabatic process in the form: $pV^k = const$, where $k = c_p/c_V$.

Investigation of properties of adiabatic process and the knowledge of Newton's and Laplace's approaches to the Torricelli's formula have allowed to deduce the formula for the outflow of the gas medium to the gas medium with the relative counterpressure $p_h/p_0$ by B.Saint-Venant and P.Wantzel to 1839:

$$V = \left( \frac{2}{k-1} kgRT \left( 1 - \left( \frac{p_h}{p_0} \right)^{\frac{k-1}{k}} \right) \right)^{\frac{1}{2}}$$

The analysis of this formula has allowed to elucidate such substantial features of gas motion, as baro-subcritical, baro-overcritical outflow and outflow into empty space. However given formula has not wished to feature flow of gas in the real flowing element, and especially in the flowing system, even in absence of heat exchange and any physical actions. The coefficients of velocity and the mass-flow rate was added to it as it is accepted in hydromechanics, adiabatic process was exchanged by the polytrope - all vainly. Only today, later 150 years, the belladonna-formula has uncovered the true sense [1].

In the first half of the XIX century the development of the elementary bases of thermodynamics happened parallel with upbuilding of knowledge about the gas laws. In 1842 J.Mayer search out the expression for mechanical equivalent of heat which is equally indispensable for generalizing the gas laws and for development of thermodynamics $c_p - c_V = R$ and 32 years also were required, when Mendeleyev has completed a history of making of gas equation including the mechanical equivalent of heat and Kelvin's temperature side by side with the special gas laws:

$$p = \frac{G}{V} RT .$$

This appearance has passed quietly enough and almost not influenced to the problem of the gas motion. At all events, the scientific association has understood that this equation result in the radical improvement in the gas dynamics neither in itself nor in combination with the fundamental Torricelli - Saint-Venant - Wantzel's formula nor in combination with the formalistic line of J.D'alembert - L.Euler - J.Lagrange - L.Navier - S.Poisson - G.Stokes and was perceived enough indifferently. And 120 years later, it turned out, that the given generalized equation requires development with the purpose of it application for description of gas stream motion.

This equation is rather productive conformably to the concrete circumstances. At the same time, it is very important whether we understand physics of processes in gas stream and significance of this equation for its description. It expressed condition of state of motionless gas medium under thermal action in a closed-loop system. As applied to open systems, for example, in the form of the flowing element (pipe), we gain again three special gas laws, having accepted the weight flow and the volume-flow rate of gas stream instead of weight and volume of motionless gas accordingly:

$$T = \frac{1}{R} \cdot \frac{Q}{\dot{G}} p = \frac{1}{R} \cdot \frac{Fw}{\dot{G}} p ; \qquad (1)$$

$$Q = \frac{\dot{G}RT}{p} \quad \text{or} \quad w = R\frac{1}{F}\dot{G}\frac{T}{p} ; \qquad (2)$$

$$p = R\frac{\dot{G}}{Q}T = R\frac{1}{F} \cdot \frac{\dot{G}}{w} \cdot T , \qquad (3)$$

where $F$ area of cross section of flowing element, $w$ - velocity of gas flow, $Q$ и $\dot{G}$ – the volume-flow rate and the mass-flow rate of flowing element accordingly. Operating by analogy with the special gas laws for the closed system we shall receive:

$$T = \frac{1}{R} \cdot \frac{Q}{\dot{G}} \cdot p = \frac{1}{R} \cdot \frac{Fw}{\dot{G}} \cdot p = const; \tag{1a}$$

$$Q = \frac{\dot{G}R}{p} T = const, \quad w = \frac{1}{F} \frac{\dot{G}R}{p} T = const; \tag{2a}$$

$$p = \frac{\dot{G}R}{Q} T = R \frac{1}{F} \cdot \frac{\dot{G}}{w} \cdot T = const \tag{3a}$$

The obtained dependences are valid for the stationary steady flows of gas medium and allow to establish the type of the process realized in gas flow, depending on the separate or combined action of heat flow and pressure on gas stream. The action by pressure in this case means its increase or decrease simultaneously and equally in spaces of the inlet and outlet surrounding the flowing element, system. According to the dependence (1a), the increase or decrease of the pressure in the above indicated spaces calls change of correlation of the volume-flow rate and the mass-flow rate, compensating the action of pressure so, that the isothermal change is ensured in gas stream. For example, the gas stream mass-flow rate in flowing element, system under action of increased pressure is incremented and the volume-flow rate (velocity) of a stream is diminished ensuring an invariability of temperature of gas in a stream, i.e. its the isothermal state.

According to the dependence (3a), the action with heat flow to gas stream is accompanied by respective change of correlation of the volume-flow rate and mass-flow rate so, that the isobaric process is ensured in a gas stream. For example, the gas stream mass-flow rate under action of heating is diminished and volume-flow rate (velocity) of gas stream is being increased ensuring an invariability of pressure in gas stream, i.e. its the isobaric state.

According to the dependence (2a), the change of mass-flow rate happens so, that the volume-flow rate (velocity) of stream is remained fixed under action simultaneously both of heat flow and pressure on gas stream. For example, the volume-flow rate and the mass-flow rate (velocity) will remain invariable under heating of stream and the simultaneous equivalent recompression in surrounding spaces of the flowing element, system. The lowering of mass-flow rate will happen if the thermal factor will predominate over the pressure, and the volume-flow rate (velocity) of stream will remain former in this case. If the factor of pressure will predominate over the thermal action, the mass-flow rate of stream will increase exactly so, as far as it is necessary that the volume-flow rate (velocity) of stream has remained invariable. In any of these cases, the analogy of state of gas stream with isochoric process in static conditions is remained for the dependence (2a).

As a whole three special laws of gas state in stream is integrated in ones expression:

$$p = \frac{\dot{G}}{Q} \cdot RT, \tag{4}$$

being dynamic analogy of the Clapeyron-Mendeleyev's equation and indispensable component for model operation of motion and state of gas stream under action of temperature and pressure on its.